\documentclass[%
superscriptaddress,
frontmatterverbose,  
preprint,
preprintnumbers,
amsmath,
amssymb, 
aps, 
12pt,
pre,
floatfix 
]{revtex4-1}

\usepackage{amsfonts}       
\usepackage{amsmath}       
\usepackage{amssymb}
\usepackage{nicefrac}

\usepackage[lofdepth,lotdepth]{subfig}

\usepackage{dcolumn}
\usepackage{bm}
\usepackage{hyperref}
\usepackage{xcolor}  
\usepackage{array}
\usepackage[export]{adjustbox}
\usepackage{multirow}
\usepackage[figure,table]{totalcount}
\usepackage{verbatim}

\usepackage{xcolor}
\usepackage{soul}

\newcommand{\avg}[1]{\left\langle#1\right\rangle} 

\begin{document}

\title{Physics of Psychophysics: two coupled square  
lattices of spiking neurons have huge dynamic range at criticality}

\author{Emilio F. Galera}

\email{emiliofgalera@gmail.com}

\affiliation{Universidade de S\~ao Paulo, FFCLRP, Departamento de F\'isica, Ribeir\~ao Preto, SP, 14040-901, Brazil}

\author{Osame Kinouchi$^*$}

\thanks{osame@ffclrp.usp.br}

\affiliation{Universidade de S\~ao Paulo, FFCLRP, Departamento de F\'isica, Ribeir\~ao Preto, SP, 14040-901, Brazil}

\date{\today}

\begin{abstract}
Psychophysics try to relate physical input magnitudes to psychological or neural
correlates. Microscopic models to account for macroscopic psychophysical laws,
in the sense of statistical physics, are an almost unexplored area. Here we
examine a sensory epithelium composed of two connected square lattices of
stochastic integrate-and-fire cells. With one square lattice we obtain a
Stevens's law $\rho \propto h^m$ with Stevens's exponent $m = 0.254$ and a
sigmoidal saturation, where $\rho$ is the neuronal network activity and $h$ is
the input intensity (external field). We relate Stevens's power law exponent
with the field critical exponent as $m = 1/\delta_h =  \beta/\sigma$. We also
show that this system pertains to the Directed Percolation (DP) universality
class (or perhaps the Compact-DP class). With stacked two layers of square
lattices, and a fraction of connectivity between the first and second layer, we
obtain at the output layer $\rho_ 2 \propto h^{m_2}$, with
$m_2 = 0.08 \approx m^2$, which corresponds to a huge dynamic range. This
enhancement of the dynamic range only occur when the layers are close to their
critical point.
\end{abstract}

\maketitle

\section{\label{sec:level1}Introduction}

Psychophysics is perhaps the oldest experimental part of Psychology, starting
with the pioneering work of Fechner in  1860~\citep{Fechner2012}. Its main aim
it to describe how sensation  is related to the input level reaching a sensory
organ. Psychophysical laws are emergent properties of neuronal
networks~\citep{Stevens1975}. A fundamental problem in these laws is that they
relate several orders of magnitude of input to few orders of magnitude of
output. This means that biological sensors have a huge dynamic range (DR) and
they should have that since natural stimuli varies by orders of intensity. For
example, several experimental results can be fitted by a Hill
curve~\citep{Chastrette1998}:

\begin{equation}\label{Hill}
S(I)  = \frac{S_\mathrm{max}\: c I^m}{S_\mathrm{max} +c I^m}\:.\,
\end{equation}

where $S(I$ is the sensation level (to me measured in some scale,
$S_\mathrm{max}$ is a saturation level, $c$ is a constant and $I$ is the input
level. For  moderate input such that $c I^m \ll S_\mathrm{max}$, we have the
famous Stevens's power
law~\citep{Stevens1957,Stevens1975,Chastrette1998,Kello2010,Teghtsoonian2012,Kornbrot2016}:

\begin{equation}
    \label{Stevens}
    S(I) = c I^m \:,
\end{equation}
where $m$ is the so called Stevens's psychophysical exponent. If $m<1$  we have
a compressive curve with large dynamic range.

It is not clear how to get this result because sensory neurons at periphery have
small dynamic range. In a sense we have a problem typical of statistical physics:
how to construct a microscopic model (here based in neurons) that explains a high
level phenomenological law as a collective phenomenon.

The idea that certain parts of the brain can benefit from operating near the
critical point of a continuous phase transition has been around for some time
now \cite{Beggs2003,Beggs2008,Chialvo2010,Shew2013,Munoz2018,Wilting2019}. In
particular,  it has been shown that criticality enables that a network of
excitable elements with small DR can present a large dynamic range as a
collective property both at the
theoretical~\cite{Kinouchi2006,Copelli2007, Assis2008,Shew2009,Larremore2011,Pei2012,Batista2014,Wang2017,Campos2017,Zhang2018}
and experimental level~\cite{Shew2009,Gautam2015, Shew2015,Antonopoulos2016}. 

A result very similar to the Hill curve is predicted by there computational
models~\cite{Kinouchi2006,Copelli2007,Assis2008}, but without a simple analytic
form as Eq.~(\ref{Hill}), suggesting that the use of a Hill curve in
Psychophysics is only a phenomenological or fitting procedure, that cannot be
obtained from first principles. This view of a large dynamic range as a
collective or emergent property (critical or not) of networks of excitable cells
is relatively new~\cite{Copelli2002,Copelli2005,Copelli2005b,Gollo2009}.

The standard textbook model to account for a large DR constructed from small DR
units is some variation of Recruitment Theory: sensory neurons, that present
sigmoidal responses with short DRs but different response thresholds, are
combined to produce a total output with a large DR. In these models, the value
of exponent $m$ is not predicted or constrained (see as
examples~\cite{Chastrette1998,clealand1999, clealand2005}). Recruitment theory
has a major flaw: for a wide range of stimulus to be perceived an equally wide
variety of receptor expression in sensory cells must occur. Experimentally,
however, receptor over-expression is only about two or
threefold~\cite{clealand1999}, so it is plausible to assume that this is not the
main mechanism responsible for the phenomenon?\cite{Copelli2007}. Also, the
model mechanism is somewhat over-simplistic: a simple linear sum of sigmoids
with a distribution of cells fitted by hand to produce a more or less
acceptable power law in some input range. 

In a sense, Recruitment Theory is a curve fitting exercise with sigmoids, not a
predictive theory. For example, it does no predicts the possible values for
Stevens's exponent $m$ (as will be done here): this exponent is only a fitting
parameter which is unable to give deep insights about the underlying neuronal
network model. On the other hand, in our theory, $m$ relates to the statistical
physics field critical exponent as $m = 1/\delta_h$.

Here we offer a modern view aboutvpsychophysical laws: any network of excitable 
cells, indeed any excitable media~\cite{Kinouchi2006,Assis2008,Copelli2002},
produces a Stevens's response $S \propto I^m$ for moderate $I$ (and a Hill's
type saturation afterwards). This is an intrinsic and basic collective property
of excitable cells. So, we need not devise some tricky mechanism to obtain
Stevens's law: it is there from start, as a basic property of any network of
excitable cells.

In this paper, we study stochastic integrate-and-fire neurons interacting in two
coupled square lattices that would be a toy model for a  biological sensor. We
assume that the coupling inside the square lattices is done by electric
synapses, as observed say in the retina or the olfactory bulb. We show that, in
each layer, there occurs a continuous phase transition from a silent phase to
an active phase. This is called an absorbing phase transition: the absorbing
phase corresponds to silence or zero activity, from what the system can not
spontaneously escape. The active phase emerges with a given critical exponent
from the critical point. Almost all of such transitions pertain to the
ubiquitous Directed Percolation (DP) class~\citep{Janssen1981,Grassberger1981},
or perhaps the so called Compact DP (C-DP or Manna) class which, in the square
lattice ($d=2$), have specific critical exponents that define its universality
class.

Our $2d$ result is new but not  surprising, since the mean field DP exponent 
$m^\mathrm{MF} = 1/\delta_h = 1/2$ has already been found for complete graphs
and random  networks~\cite{Kinouchi2006,Brochini2016,Girardi2020}. However, our
$2d$ DP exponent $m \approx 0.254$ means that $S \propto I^{0.254}$, being a
huge improvement over the mean-field $S \propto I^{0.5}$ result.

The second square lattice layer is put after the first one, with a small
fraction $p$ of electrical synapses between them. The aim of constructing such
retina-like two-layered sensor is to show that other Stevens's exponents can be
obtained by changing the network topology.  This has already been demonstrated
for random networks~\cite{Martinez2020}, where the second layer presented
Stevens's exponent $m_2^\mathrm{MF} = 1/4 = (m^\mathrm{MF})^2$ (the index in
$m_2$ refers to the second layer output). Here we obtain a similar result, but
with a huge dynamic range given by  $m_2 = 0.078 \approx m^2$, that is, an input
range of $O(10^{12})$ (similar to the difference between luminosity at midnight
and noon) can be mapped onto an $O(1)$ output.

\section{\label{sec:level2} Model and Methods}

We use a stochastic leaky integrate-an-fire neuron originally proposed by
Gerstner and van Hemmen~\cite{Gerstner1992}, rigorously investigated by Galves
and Löcherbach~\cite{Galves2013}, Ferrari {\em et al.}~\cite{Ferrari2018} and 
simplified in~\cite{Brochini2016, Costa2017,Kinouchi2019}. Time is discrete and
updates are done in parallel. For a discussion about LIF discrete time models, 
see~\cite{Cessac2008,Cessac2010,Cessac2011}.

The  membrane potential of a neuron situated at the $i$-th line and $j$-th
column is given by:

\begin{eqnarray}
    \label{eq:mod2D}
    V_{ij}[t + 1] = \mu V_{ij}[t] + I_{ij}[t] + \frac{1}{4}
    \sum^4_{kl \in \cal{V} } W_{ij,kl}X_{kl}[t]\:,  &\iff& X_{ij}[t] = 0 \:,\\
    \label{eq:reset2D}
    V_{ij}[t + 1] = 0\:,   &\iff& X_{ij}[t] = 1 \:.
\end{eqnarray}

Here, $X_{ij}[t]$ is a binary state variable ($X=1$ = spike, $X = 0$ = silence).
A neuron stays in its active state for only one time step, assumed here as the
typical spike width of $1$ ms. If at a given time $t$ the neuron with indexes
$ij$ fires, its membrane potential is reset to zero,  Eq.~(\ref{eq:reset2D}),
otherwise the neuron will follow Eq.~(\ref{eq:mod2D}). The parameter
$\mu \in [0, 1]$ is a leakage term which controls how much the neuron remembers
from its previous voltage $V_{ij}[t]$, $I_{ij}[t]$ is the external input.
Neurons interact in the two dimensional lattice where each one is connected to
its four nearest neighbors (the von Neumann four sites neighborhood is
$\cal{V}$.). The strength of the (electric) synapse between the postsynaptic
neuron $ij$ and the presynaptic neuron $kl$ (with $k = i$ and $l = j \pm 1$ or 
$k = i \pm 1$ and $l =j $) is denoted  by $W_{ij,kl}$. We use periodic boundary
conditions.

In this stochastic model, the firing of a neuron is probabilistic and given by a
firing function $\Phi$:

\begin{equation}
    \label{eq:probfire}
    P\big( X_{ij}[t] = 1 | X_{ij}[t - 1] = 0 \big) = \Phi(V_{ij}[t]) \:.
\end{equation}

Notice that this form emphasizes that the neuron has one time step of  absolute
refractory period, although this is implicit because we assume $\Phi(V=0) = 0$.
The function $\Phi$ needs only to be a sigmoidal function. For mathematical
convenience, we use the so called ``rational
function''~\cite{Costa2017,Kinouchi2019} :

\begin{equation}
    \label{eq:phi}
    \Phi(V) = \frac{\Gamma(V - \theta)}{1 + \Gamma(V - \theta)}\:\Theta(V - \theta) \:.
\end{equation}

Here,  $\theta$ is a firing threshold bellow which the neuron cannot fire, i.e
$\Phi(V) = 0$ for $V < \theta$. The $\Gamma$ parameter in Eq.~(\ref{eq:phi}) is
the neuronal gain. Notice that both the gain $\Gamma$ and thresold $\theta$ are
parameters experimentally related to the phenomenon of firing rate adaptation
\cite{Benda2003, Buonocore2016}. Although not implemented here, a homeostatic
dynamics in the individual synapses $W_{ik}$ and thresholds $\theta_{ij}$ can be
used to self-organize the network towards the critical state~\cite{Girardi2020}.
So, in our discussion about maximizing the dynamic range at criticality, we will
assume that such homeostatic mechanisms can operate in our network.

For a meticulous analysis of the mean-field approximation regarding this model
we refer to \cite{Brochini2016, Costa2017,Kinouchi2019}. We will show that the
system undergoes an absorbing second order phase transition if the external
field and the firing thresholds are  equal, $\theta_{ij} = I_{ij}$,
see~\cite{Girardi2020}. This apparent fine tuning  condition will be discussed
in Sec.~\ref{H}. If $\theta_{ij} \neq I_{ij}$, the transition is of first
order~\cite{Brochini2016,Costa2017}.

The activity of a network with $N$ neurons is, at any time,
\begin{equation}
    \label{eq:rho}
    \rho[t] = \frac{1}{N}\sum_{ij}X_{ij}[t],
\end{equation}
it can be measured by
\begin{equation}
    \label{eq:rhost}
    \rho(w,\Gamma) = \frac{1}{t_f - t_i}\sum_{t = t_i}^{t_f}\rho[t],
\end{equation}
where $(t_i, t_f)$ marks a large time interval in the simulation far from
transient states. The time average $\rho(W,\Gamma)$, Eq.~(\ref{eq:rhost}), is
used as our order parameter and the average synaptic weight
$W = \langle W_{ij,kl }\rangle$ and average gain
$\Gamma = \langle \Gamma_{ij }\rangle$ are our control parameters. We notice
that, if at a given time step, we have $\rho[t] = 0$, then a random site is
chosen and its state is put as $X_i[t+1]=1$. This is done to get the network out
of the absorbing state.

We define another  quantity, the order parameter fluctuations $\Delta \rho$,
related to the fluctuations of the activity $\rho$,
\begin{equation}
    \label{eq:chi}
    \Delta \rho = N\big(\langle \rho^2 \rangle - \langle \rho \rangle^2\big) \:.
\end{equation}

In order to evaluate how the neuron model responds to external stimulus when
interacting in a layered system one needs to know how a single $2d$ lattice of
neurons work. First, with $\Gamma =1$ fixed, we explore the effect of the
control parameter $W$ to determine roughly where the phase transition occurs,
i.e, we need to know where are the sub-critical, critical and super-critical
regimes. Then, we can refine our measurements and explore the critical region
for systems with different sizes $N$ and use finite-size scaling techniques
\cite{Hinrichsen2000} to determine better the critical point of the transition
$W_c$, the order parameter critical exponent $\beta$  and the  susceptibility
critical exponent $\gamma^\prime$.

In the vicinity of the critical point, $\rho$ and $\Delta \rho$ should scale as,
\begin{equation}
    \label{eq:rhocrit}
    \rho \propto |\overline{W}|^{\beta}\:, \:\:\:\:\:\:\:\:
    \Delta \rho \propto |\overline{W}|^{-\gamma^\prime} \:,
\end{equation}
where $\overline{W}$ is the reduced control parameter
$\overline{W} = (W - W_c)/W_c$.

We then proceed to our \textit{bottom-up} psychophysics approach, where we build
a sensor with two layers of neurons. Each layer is a square lattice. The first
one is stimulated by the external stimulus, modeled as a Poisson process. The
probability  per time step of a neuron being activated is:

\begin{equation}
    \label{eq:poiss}
    \lambda = 1 - e^{-r},
\end{equation}

Here, $r$ is a stimulation rate. Neurons in the first layer can either fire due
to synaptic excitation from neighbors or with probability $\lambda$ due to the
input Poisson process.

On the second layer, we chose randomly $p = 0.1 N$ neurons to connect to the
first layer. These neurons receive an input  $I_{ij}[t] = J X_{ij}^1[t]$ where
$X_{ij}^1$ is the neuron of the first layer, that is, both connected neurons of
the first and second layer share the same indexes $ij$. The value $J$ used was
high enough to guarantee that the connection between layers always pass
information, that is, if a neuron in the first layer spikes, then, the neuron
connected to it in the second layer also forcibly spikes. The electric synapses
$J$  are unidirectional, that is, activity in the second layer does not excites 
back the first layer.

We ran simulations for various system sizes in the three regimes, always
stimulating neurons from the first layer with rates ranging from $r = 10^{-6}$
up to $r = 0.1$. We will show that in, the critical region, the activity $\rho$
in each layer can map stimulation rates $r$ with large dynamic range. For the
first layer, 
\begin{equation}
    \label{eq:rhorl1}
    \rho_{1}(r) \propto r^{m}\:,
\end{equation}
and for the second layer,
\begin{equation}
    \label{eq:rhorl2}
    \rho_{2}(r) \propto (\rho_1(r))^{m} \propto r^{m^2}.
\end{equation}

Critical systems in the presence of an external field $h$ have a well
established behavior. For small field, the order parameter scales as a power
law,
\begin{equation}
    \label{eq:rhome}
    \rho(h) \propto h^{\beta/\sigma},
\end{equation}
where $\beta$ is the order parameter critical exponent and $\sigma$ is the
critical exponent associated with the mean cluster size. If we identify the
stimulation rate $r$ as the external field $h$, it is possible to write
Stevens's exponent $m$ of Eq.~(\ref{eq:rhorl1})  as:

\begin{equation}
    \label{eq:meq}
    m = \frac{\beta}{\sigma}.
\end{equation}

Eq.~(\ref{eq:rhome}) is valid for asymptotically small fields,
$h \rightarrow 0$. This means that the relation between exponents (\ref{eq:meq})
is valid as long as the stimulation rate $r$ of the Poisson process
(\ref{eq:poiss}) is small.

To quantify how each layer responds to the stimulation, we follow the dynamic
range definition of Kinouchi and Copelli~\cite{Kinouchi2006}.  We measure the
individual activity of each layer as a function of the stimulus rate $r$. We
then calculate the dynamic range $\Delta_h$ of each layer for the three regimes:
\begin{equation}
    \label{eq:dr}
    \Delta_h = 10 \log \Big( \frac{r_{0.9}}{r_{0.1}} \Big),
\end{equation}
where the values $r_{0.9}$ and $r_{0.1}$ are the stimulation rates which evoke
the activities $\rho_{0.9}$ and $\rho_{0.1}$. These activity values are obtained
through the equation $\rho_{x} = \rho_{0} + x(\rho_{max} - \rho_{0})$, where the
values $\rho_{max}$ and $\rho_{0}$ are just the largest and smallest (not
necessarily zero, due to self-sustained supercritical activity) response of a
layer. The dynamic range is a measure that relates, in decibels, the largest and
smallest inputs that the system can map in the output.

\section{\label{sec:level3} Results}

\subsection{\label{sec:level3A} The $2d$ lattice with leakage $\mu = 0$}

\begin{figure}[hbt!]
    \centering
    \subfloat[][]{
        \includegraphics[width=0.49\linewidth]{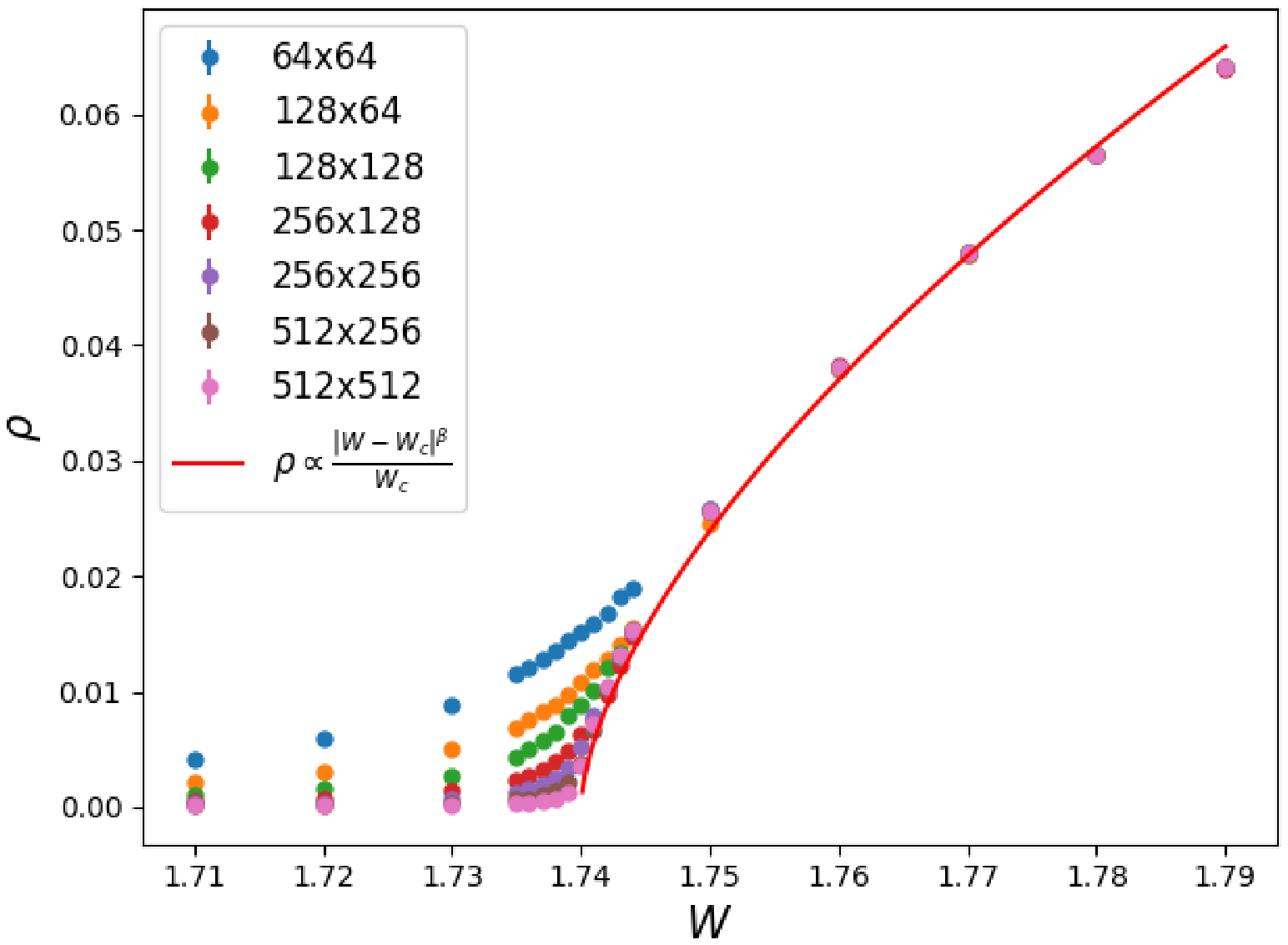}
        \label{fig:chirhoxwA}
    }
    \subfloat[][]{
        \includegraphics[width=0.49\linewidth]{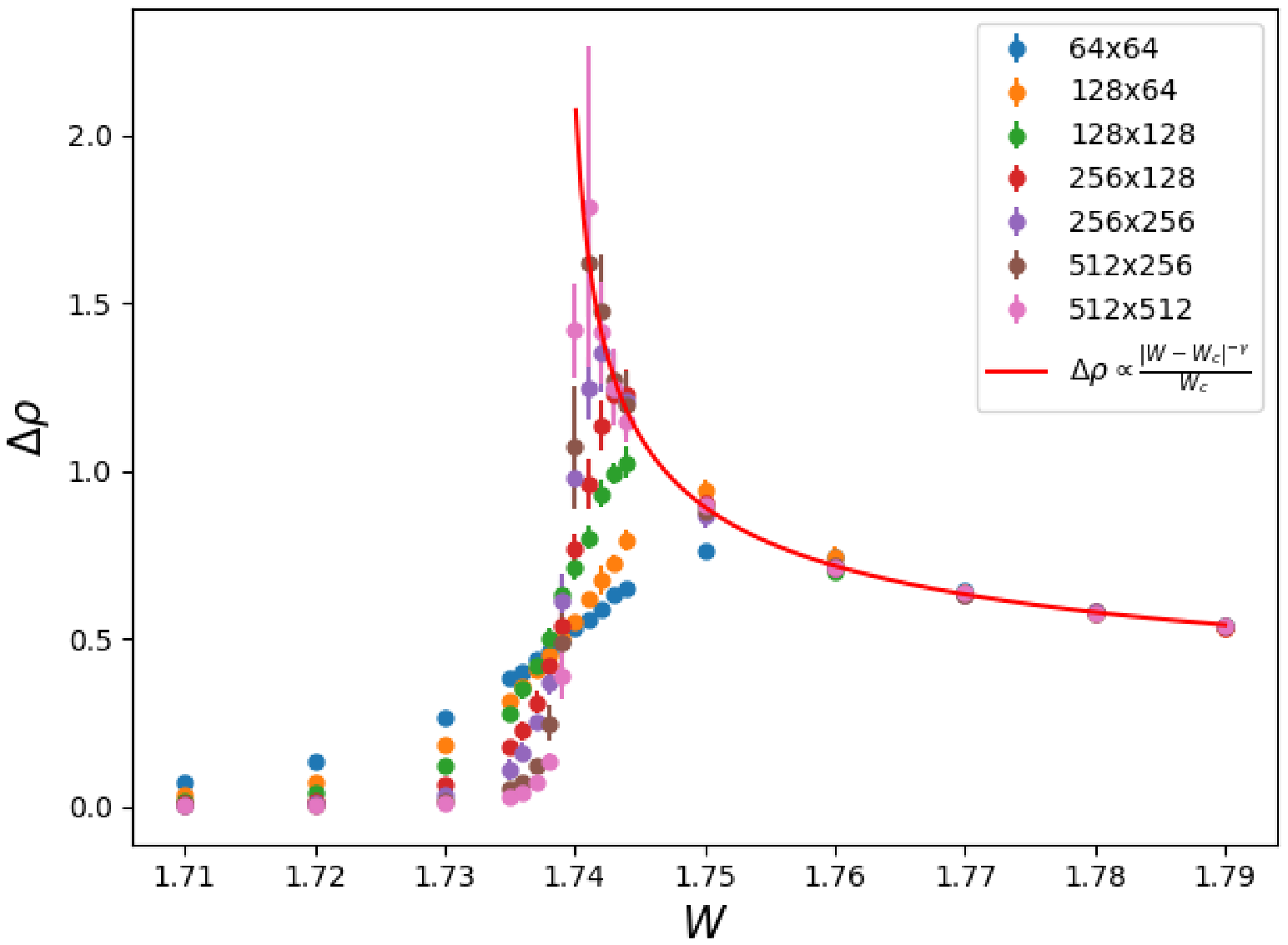}
        \label{fig:chirhoxwB}
    }
    \caption{a) Order parameter $\rho(W;N)$ for systems of different sizes close to the critical region. b) Order parameter fluctuations $\Delta \rho(W;N)$ for various system sizes near the critical region.}
    \label{fig:chirhoxw}
\end{figure}

For the $2d$ network, we first present curves $\rho(W;N)$ for different square
and rectangular lattices (from here we fix $\Gamma =1$ without loss of
generality), see~Fig.~\ref{fig:chirhoxwA}.

The solid line  has a form:

\begin{equation}\label{rho}
\rho(W;N\rightarrow\infty) \propto \overline{W}^\beta \:,   
\end{equation}
with the $2d$ critical point $W_c = 1.74$. We can see that this curve produces a
very good fit of the data for large $N$ if we use the tabulated $2d$ DP critical
exponent  $\beta = 0.583$~\cite{Lubeck2004}. We also plot  the  fluctuations
$\Delta \rho(W;N)$ as a function of the control parameter $W$ in
Fig.~\ref{fig:chirhoxw}b. We obtain a good fit
$\Delta \rho \propto \overline{W}^{-\gamma^\prime}$  for large networks if we
use the 2d DP tabulated exponent $\gamma^\prime = 0.2998$\cite{Lubeck2004}.

The significance of the DP transition for our neuronal network model is the
following. The model admits a silent phase ($\rho = 0$) but, with increasing
coupling, there occurs a change of phases. This could be an oscillatory phase or
a coexistence with two fixed points for the same  coupling (a non trivial $\rho$
phase and the trivial $\rho^0$ absorbing phase). This last case is achieved by a
discontinuous (firs order) phase transition. 

In our case, we found a continuous (second order) transition from the absorbing
state $\rho^0$ to an active phase $\rho$ at a critical point $W_c$. As discussed
in the method section, critical exponents can be defined only for continuous
transitions. The found exponents  enable us to classify our transition as
pertaining to the Directed Percolation (DP) universality class, or perhaps the
Compact DP (Manna) class. This is not so surprising, because almost all
continuous transitions from a single absorbing state pertain to such clases  and
follows the so called Janssen-Grassberger
conjecture~\cite{Janssen1981,Grassberger1981}. What perhaps is a bit curious is
that such conjecture works in a model with somewhat complicated elements like
our stochastic LIF neurons.

\begin{figure}[hbt!]
    \centering
    \subfloat[][]{
        \includegraphics[width=0.49\linewidth]{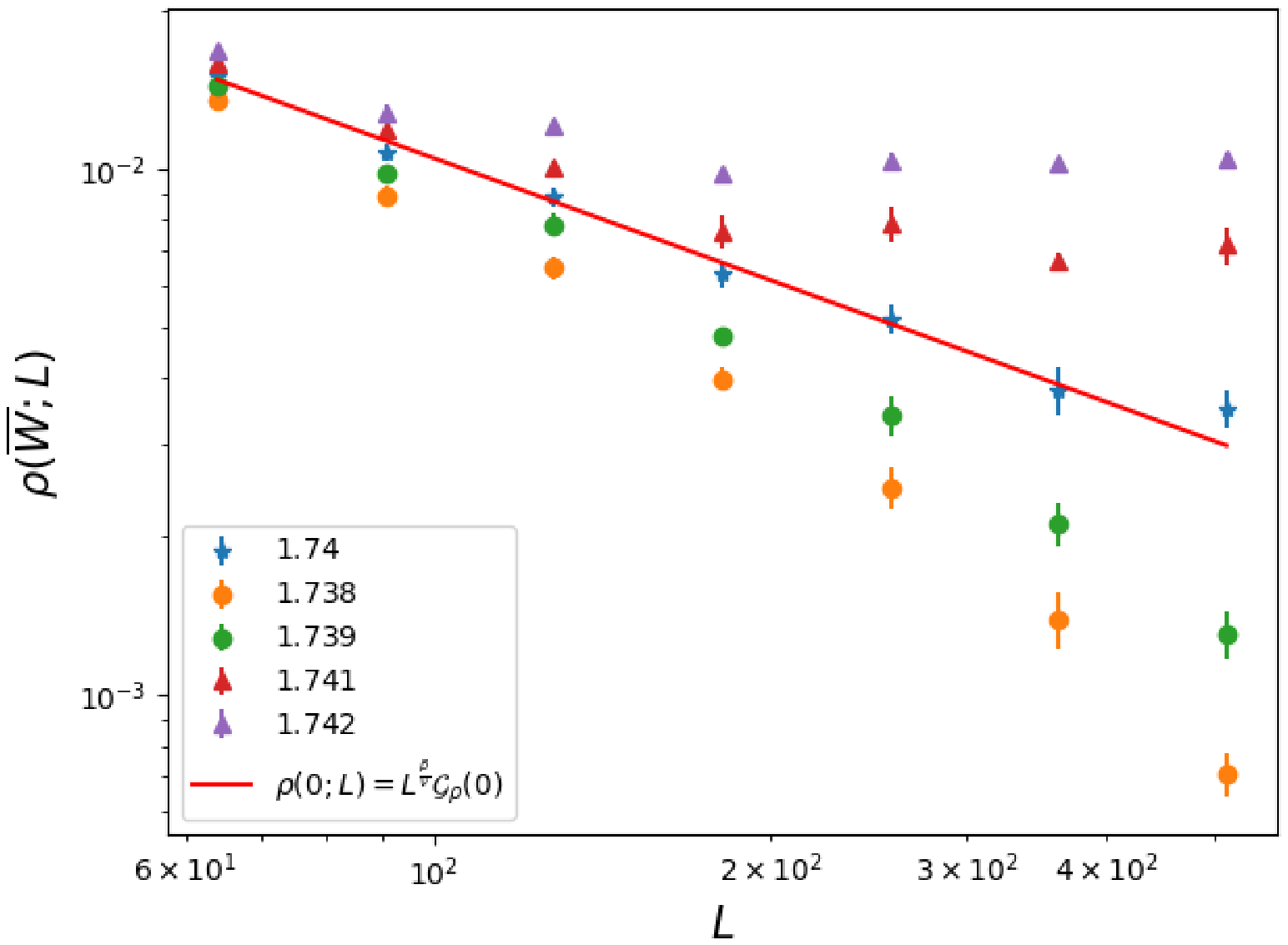}
        \label{fig:fssrho-chiA}
    }
    \subfloat[][]{
        \includegraphics[width=0.49\linewidth]{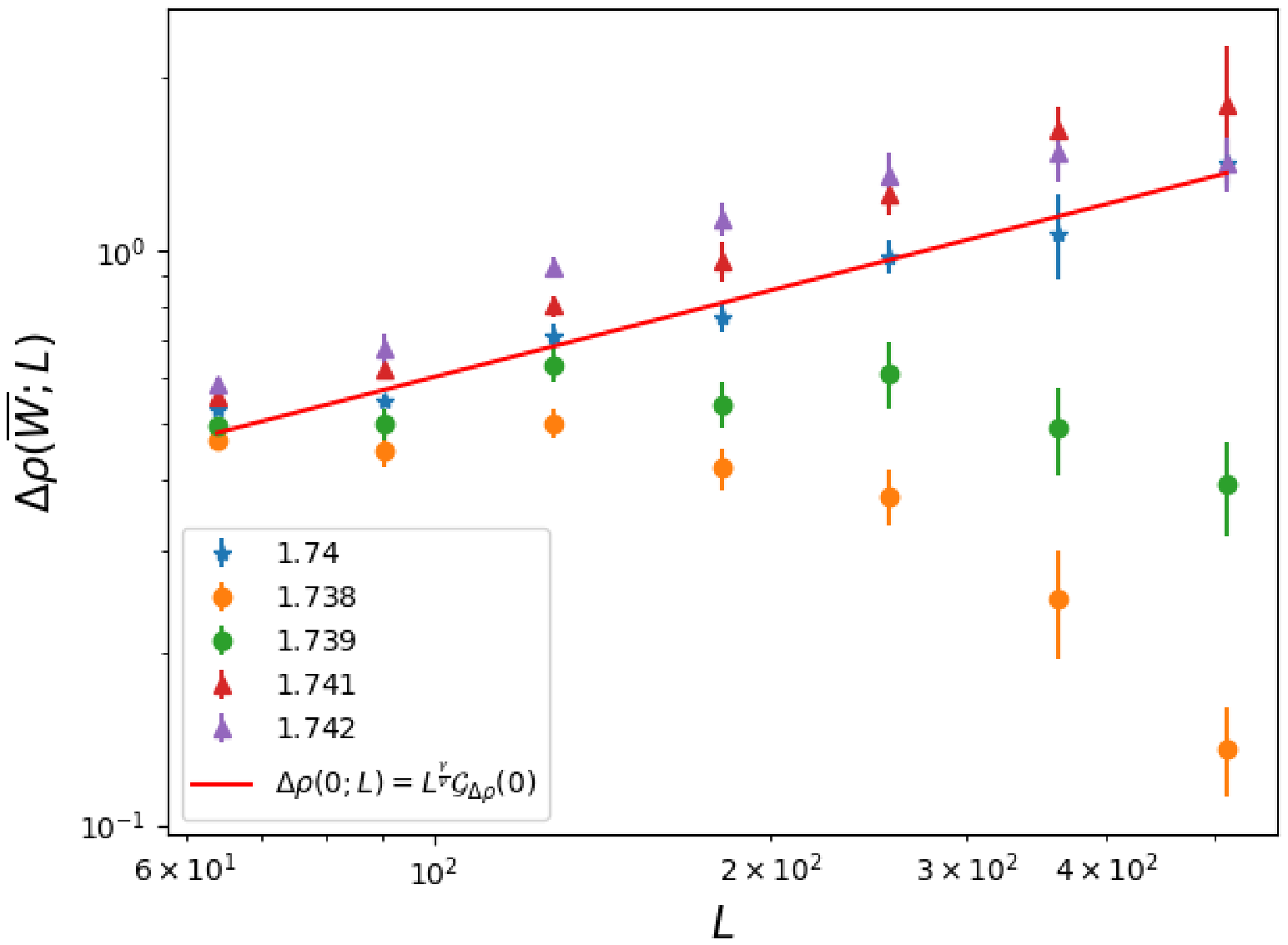}
        \label{fig:fssrho-chiB}
    }

    \caption{a) Order parameter $\rho$ and b) $\Delta \rho$ as function of $L$.
    Inverted triangles mark the critical point $W_c=1.74$. Red lines are curve
    fits to measure the exponents ratios of Eq.~(\ref{eq:fss-rx}):
    $\beta/\nu_\perp \approx 0.77$ and $\gamma'/\nu_\perp \approx 0.50$.}
    \label{fig:fssrho-chi}
\end{figure}

We tried, however, to obtain the critical exponents  in an independent way. In
the critical region, both $\rho_N$ and $\Delta \rho_N$ are strongly dependent on
the system size $N$. To get around the fact that we are far from the
thermodynamic limit, we use  \textit{finite-size scaling} techniques. First, we
re-scale $\rho$ and $\Delta \rho$:

\begin{eqnarray}
    \label{eq:fss}
        \rho(\overline{W}; L) &=& L^{-\beta/\nu_{\perp}} G_{\rho}(L^{1/\nu_{\perp}}|\overline{W}|) \:,\\
        \Delta \rho(\overline{W}; L) &=& L^{\gamma^\prime/\nu_{\perp}} G_{\Delta \rho}(L^{1/\nu_{\perp}}|\overline{W}|)\:,
\end{eqnarray}

where  $L = \sqrt{N}$ is the characteristic square network size, $\nu_{\perp}$
is the spatial correlation length critical exponent and $G_{\rho}$ and
$G_{\Delta \rho}$ are scaling functions. Then, we plot $\rho$ and $\Delta \rho$
as functions of the characteristic system size $L$ for different values of $W$.
When $W = W_c$, the reduced control parameter is $\overline{W} = 0$. The
re-scaled parameters $\rho$ and $\Delta \rho$ are then power laws which depend
only on $L$,
\begin{eqnarray}
    \label{eq:fss-rx}
        \rho(0; L) &=& L^{-\beta/\nu_{\perp}} G_{\rho}(0) \:,\\
        \Delta \rho(0; L) &=& L^{\gamma^\prime/\nu_{\perp}}G_{\Delta \rho}(0)\:.
\end{eqnarray}

The fine determination of $W_c, \beta, \gamma^\prime$ and $\nu_{\perp}$ can be
done with standard finite size techniques, as delineated above. However, this is
not our main concern here since that would require much more computational
effort and is not the main subject of the paper. By now, it is sufficient to
show that the critical exponents of our stochastic integrate-and-fire neuronal
network are compatible with the 2d DP class or with the 2d Compact-DP class
(Manna class), see Table I.

\begin{table}[htbp]
\centering
\caption{Critical Exponents (Directed Percolation and Manna exponents from~\cite{Lubeck2004}).}
\label{table1}
\resizebox{0.5\textwidth}{!}{\begin{tabular}{|l|l|l|l|l|}
\hline
2d Exponent& \multicolumn{3}{c|}{} \\  \hline
& Results  &  DP  & Manna \\ \hline
$\beta$ & $ 0.56 \pm 0.05 $  & $ 0.5834 \pm 0.0030 $  & $ 0.624 \pm 0.029$  \\ \hline
$\nu_\perp $ & $ 0.73 \pm 0.05 $   & $ 0.7333 \pm 0.0075 $  & $ 0.799 \pm 0.014 $ \\ \hline
$\gamma^\prime $ & $0.37 \pm 0.05 $  & $ 0.2998 \pm 0.0162$  & $ 0.367 \pm 0.019$  \\ \hline
$m = 1/\delta_h $ & $ 0.25 \pm 0.05 $  & $ 0.268 \pm 0.001 $ & $ 0.280 \pm 0.014$ \\ \hline
$\sigma = \beta/m $  &  $2.22 \pm 0.05 $ & $ 2.1782 \pm 0.0171 $ & $ 2.229 \pm 0.032 $  \\ \hline
\end{tabular}}
\end{table}

First, simulation results for large $N$, varying $W$, indicates
$W_c \rightarrow 1.74 \pm 0.01$. Then, a simple fit of the data in
Fig.~\ref{fig:chirhoxw}a to Eq.~(\ref{eq:rhocrit}) yields
$\beta \approx 0.564 \pm 0.05$. By using Eqs.~(\ref{eq:fss-rx}) and
Fig.~\ref{fig:fssrho-chi} fits, we obtain $\beta / \nu_\perp \approx 0.77$  and
$\gamma^\prime / \nu_\perp = 0.50$, that is,
$\nu_{\perp} \approx 0.73 \pm 0.06$ and $\gamma^\prime \approx 0.37 \pm 0.04$,
see Table~\ref{table1}. We also have obtained, from the dependence on the
external field at the critical point, $m = 1/\delta_h = 0.254 \pm 0.005$, see
Sec.~\ref{sec:level3B}. From the equality
$\delta_h = \sigma/\beta$~\cite{Lubeck2004} we obtain $\sigma = 2.22 \pm 0.05 $
We observe that the errors are not statistical but simple fitting errors, see
Table I.

\begin{figure}[ht]
    \centering
    \subfloat[][]{
        \includegraphics[width=0.49\linewidth]{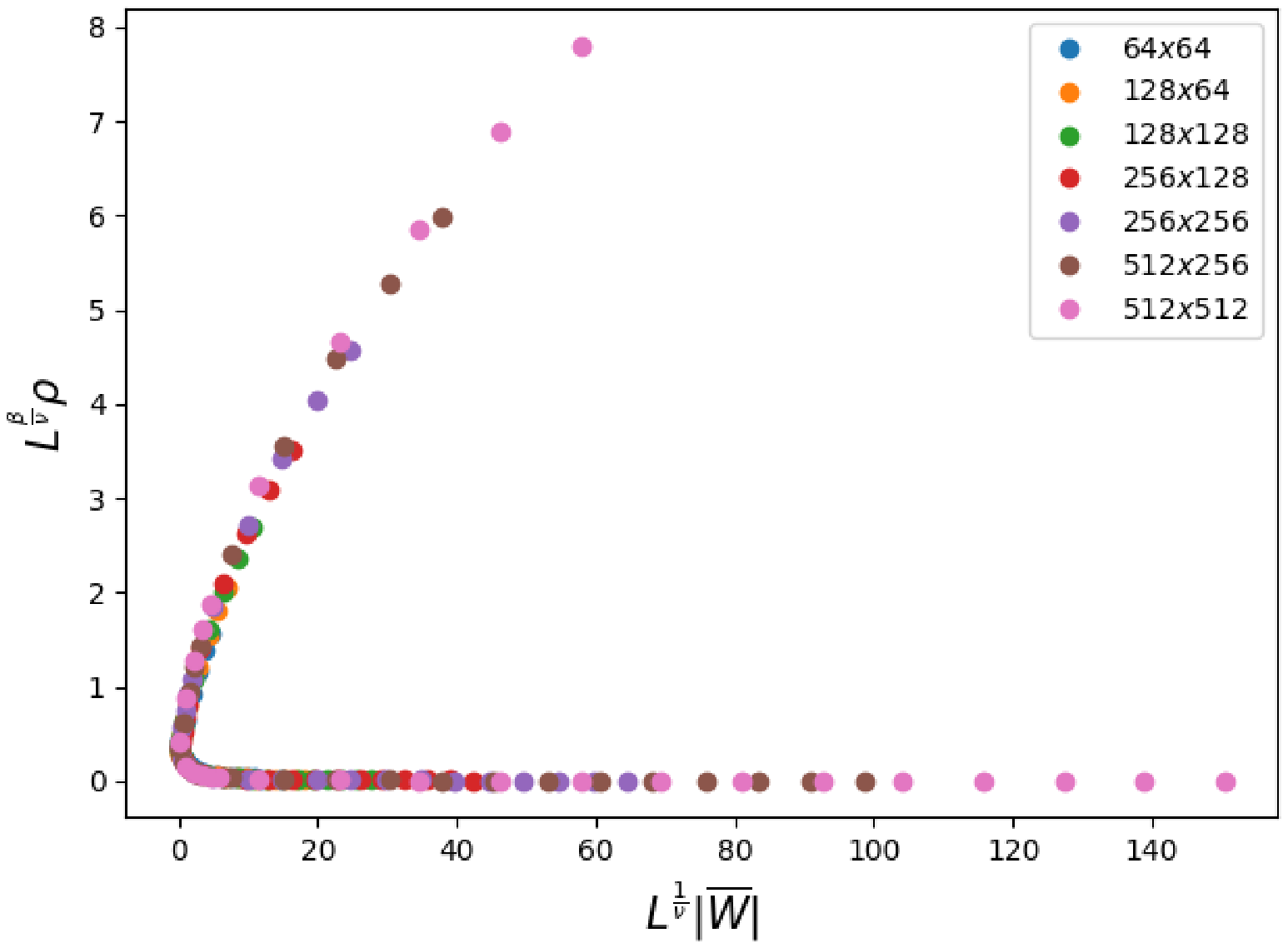}
        \label{fig:cllpsrho-chiA}
    }
    \subfloat[][]{
        \includegraphics[width=0.49\linewidth]{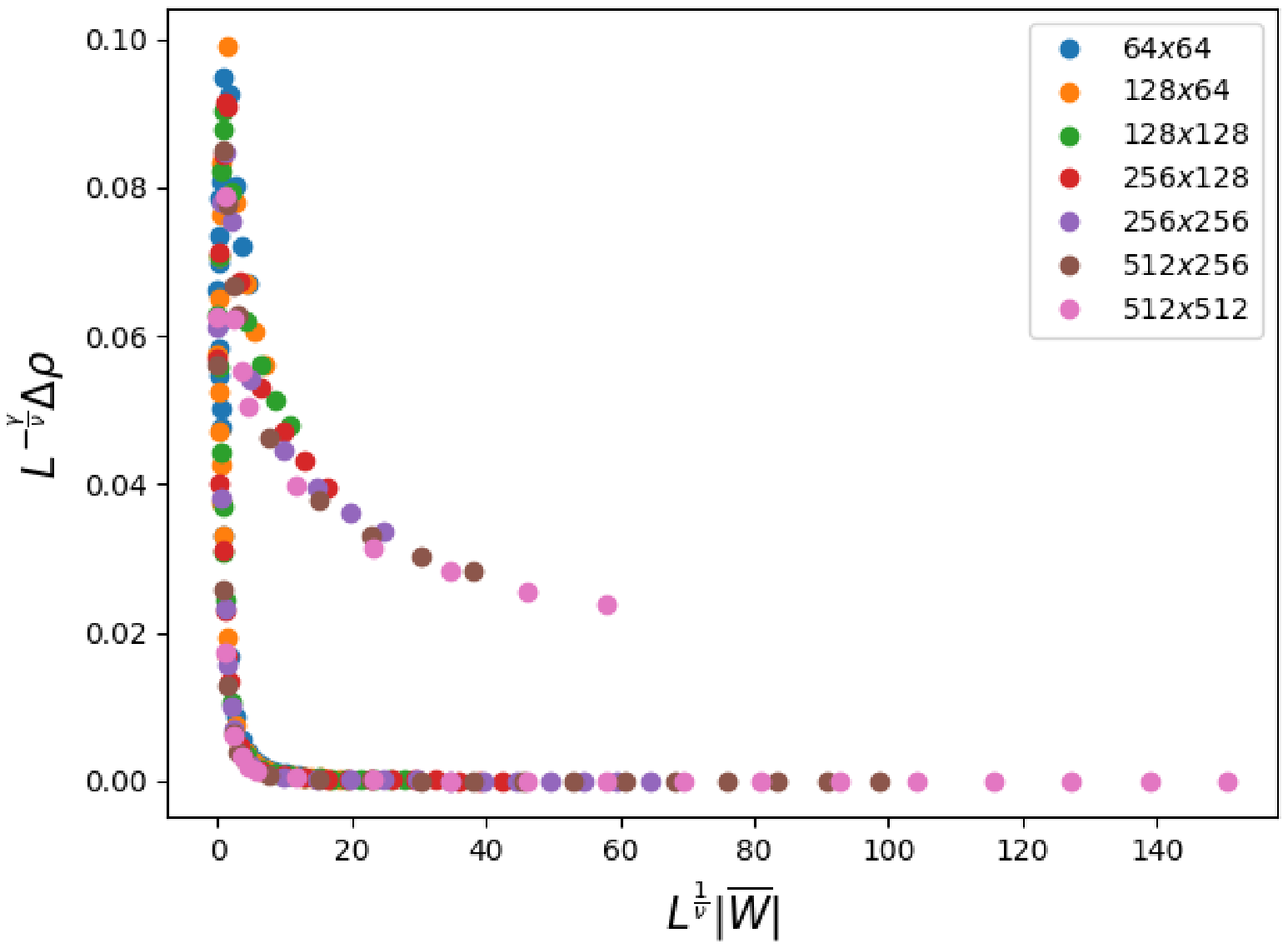}
        \label{fig:cllpsrho-chiB}
    }
    \caption{a) Data collapse for the order parameter $L^{\beta/\nu_\perp} \rho(\overline{W})$ and b) fluctuations $L^{-\gamma^\prime/\nu_\perp} \Delta \rho$.}
    \label{fig:cllpsrho-chi}
\end{figure}

We also present a data collapse by plotting curves
$L^{\beta/\nu}\rho \times L^{1/\nu}|\overline{W}|$ and
$L^{-\gamma/\nu}\Delta \rho \times L^{1/\nu}|\overline{W}|$, see
Fig.~\ref{fig:cllpsrho-chi}. Beside showing an agreement of the exponents, the
collapse gives the scaling functions $G_{\rho}$ and $G_{\Delta \rho}$. The
obtained critical exponents, the data collapse of the
$\rho_N(\overline{W})$ and $\Delta \rho_{N}(\overline{W})$ curves and the
Janssen \&  Grassberger conjecture~\cite{Janssen1981, Grassberger1981} seems to
be enough reasons to consider  that our stochastic leaky integrate-and-fire
neuronal network indeed belongs to the DP or the C-DP universality classes (the
exponents of these two classes are very similar and it is difficult to determine
the class only from the numerical results, see Table I). 

\subsection{The case with leakage $\mu > 0$}

The leakage parameter $\mu$ in Eq.~(\ref{eq:mod2D}) is the ingredient that makes
our neuron different from a simple binary automaton and to be defined as an
integrate-and-fire element. The neuron has memory of its previous inputs because
forgets its membrane potential with a time scale given by $\mu$.

\begin{figure}[hbt!]
    \centering
    \subfloat[][]{
        \includegraphics[width=0.49\linewidth]{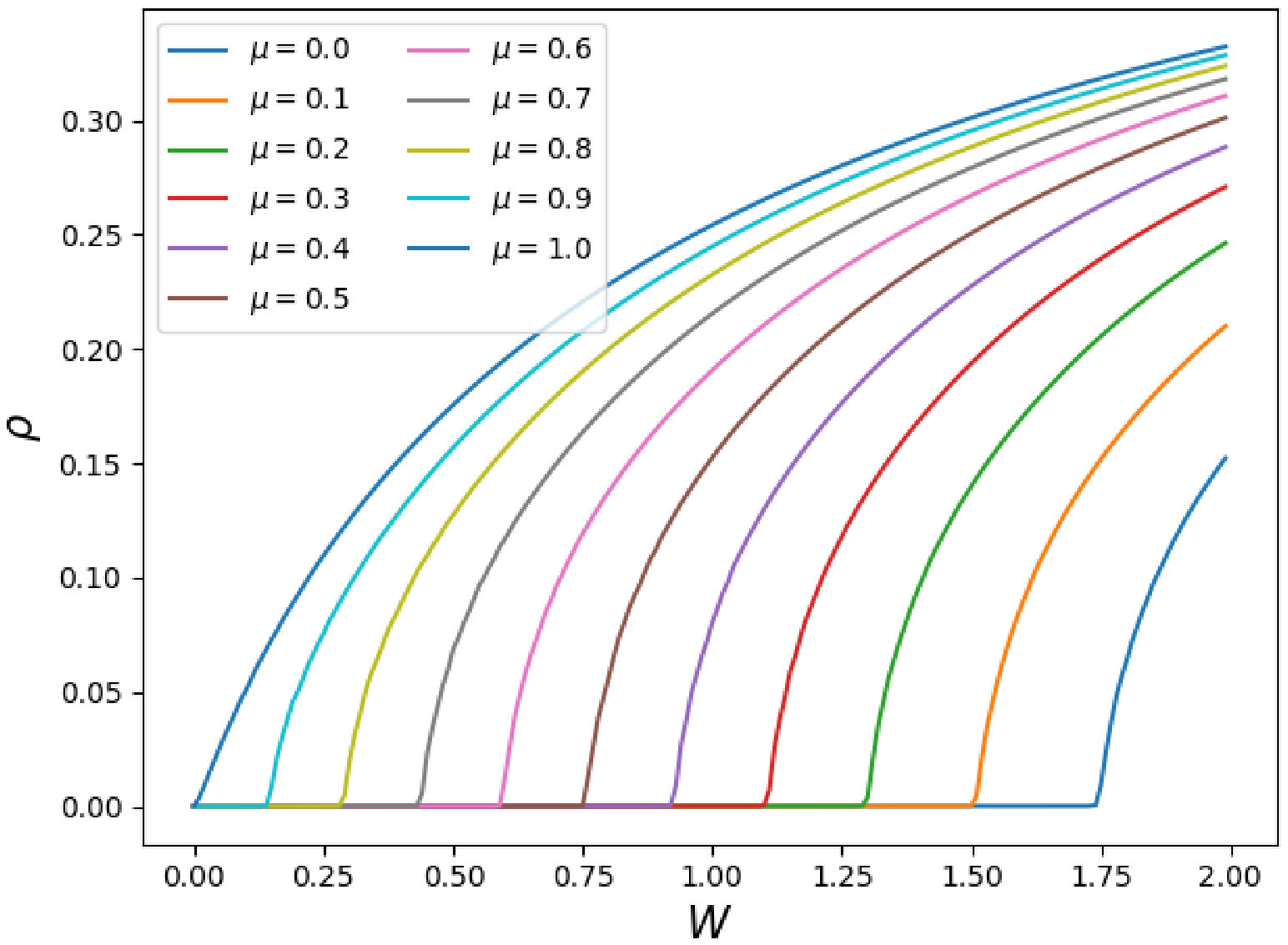}
        \label{fig:muA}
    }
    \subfloat[][]{
        \includegraphics[width=0.49\linewidth]{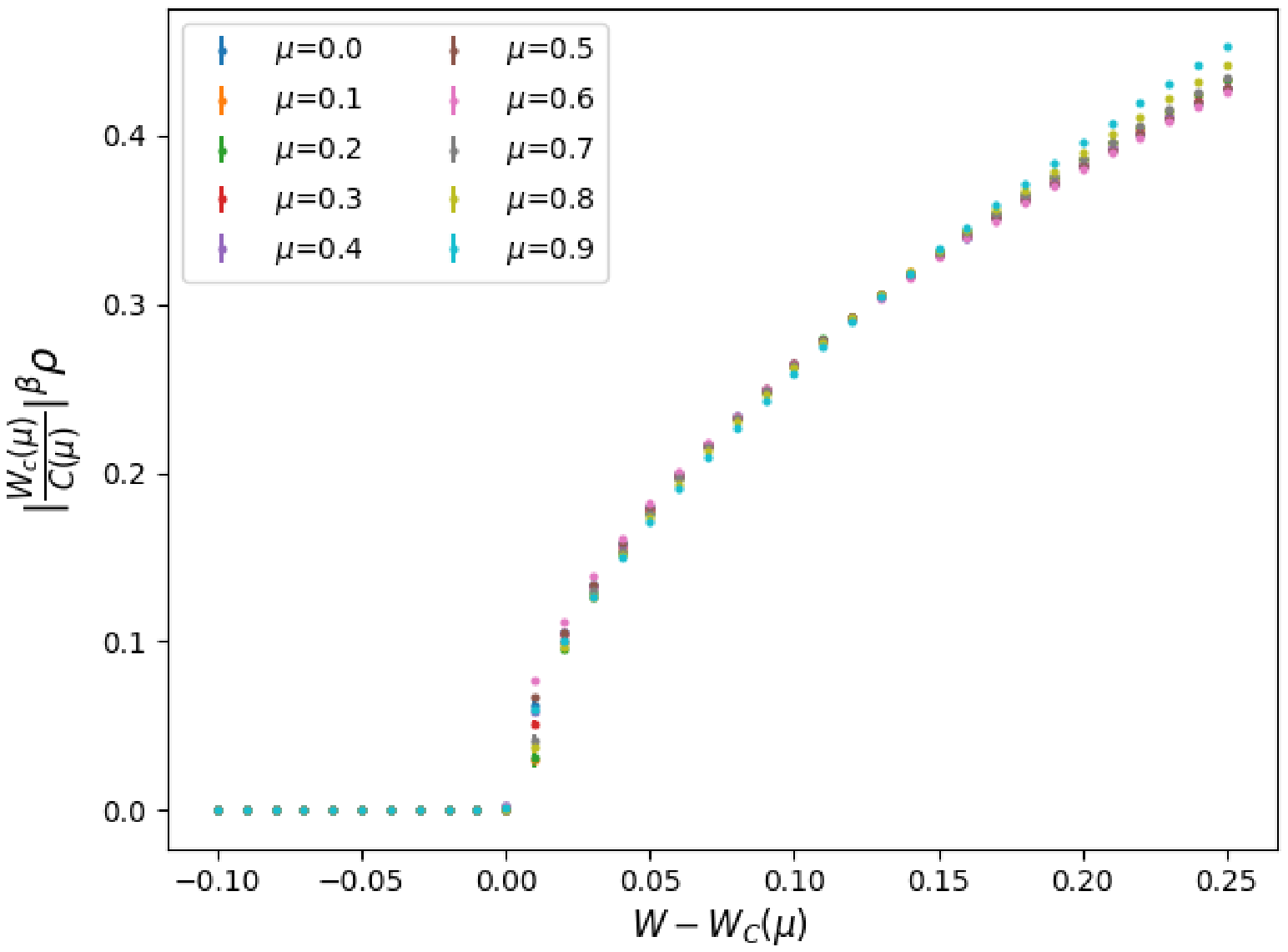}
        \label{fig:muB}
    }

    \caption{a) Curves $\rho(W,\mu)$ for different $\mu$ and
    $N = 512 \times 512$; b) Data collapse using $x = W - W_c(\mu)$ and
    $y = [W_c(\mu)/C(\mu)]^\beta \rho(\mu)$;}
    \label{fig:muAB}
\end{figure}

We found that $\mu > 0$ changes the location of $W_c$ (Fig.~\ref{fig:muA}). In
the mean-field case,  we obtain analytically that
$W_c(\mu) = (1-\mu) W_c(0)$, with $W_c(0) = 1$
(for  $\Gamma = 1$)~\cite{Brochini2016,Kinouchi2019}:

\begin{equation}
\rho(\mu) = \left[C(\mu) \frac{W - W_c(\mu)}{W_c(\mu)}\right]^\beta\:,    
\end{equation}

where $C(\mu)$ is independent of $W$. This means that, if $x = W - W_c(\mu)$,
the function $y(x) = [W_c(\mu)/C(\mu)]^\beta \rho(\mu)$ is independent of $\mu$.

We searched for a similar result for our square lattice, but now with
$W_c(\mu) = (1-\mu) W_c(0) = (1-\mu) 1.74$. We find that the collapse is not
good (not show), meaning that the mean-field result $W_c(\mu) = (1-\mu)W_c(0)$
does not generalizes to the 2d case. However, by using the measured 2d
$W_c(\mu)$ from Fig.~\ref{fig:muA}, we obtain a very good  data collapse, see
Fig.~\ref{fig:muB}. This collapsed data means that leakage $\mu > 0$ does not
change the universality class of our system.

\subsection{\label{sec:level3B} Dynamic range of the first layer}

\begin{figure}[hbt!]
    \centering
    \subfloat[][]{
        \includegraphics[width=0.49\linewidth]{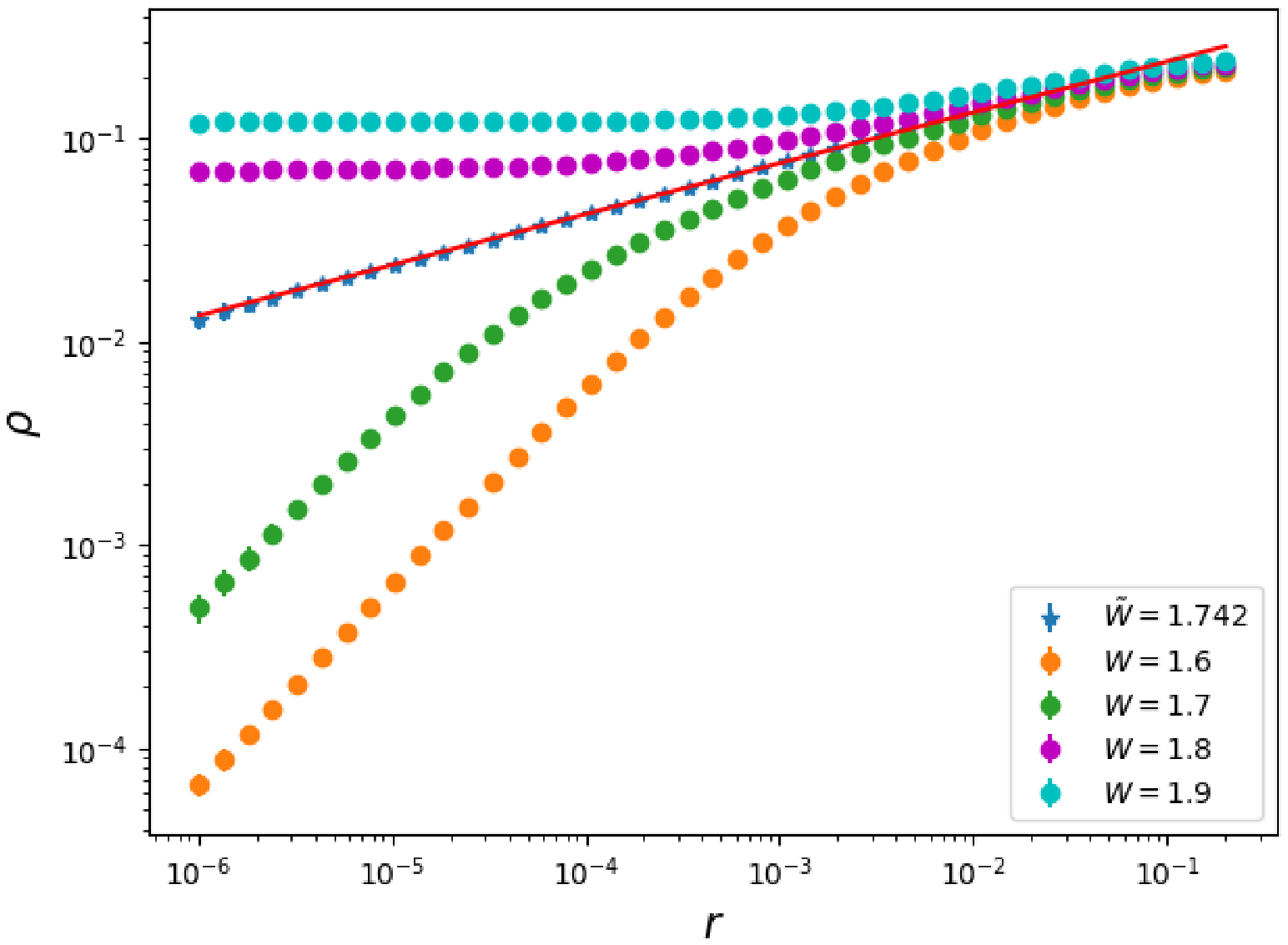}
        \label{fig:respfl1}
    }
    \subfloat[][]{
        \includegraphics[width=0.49\linewidth]{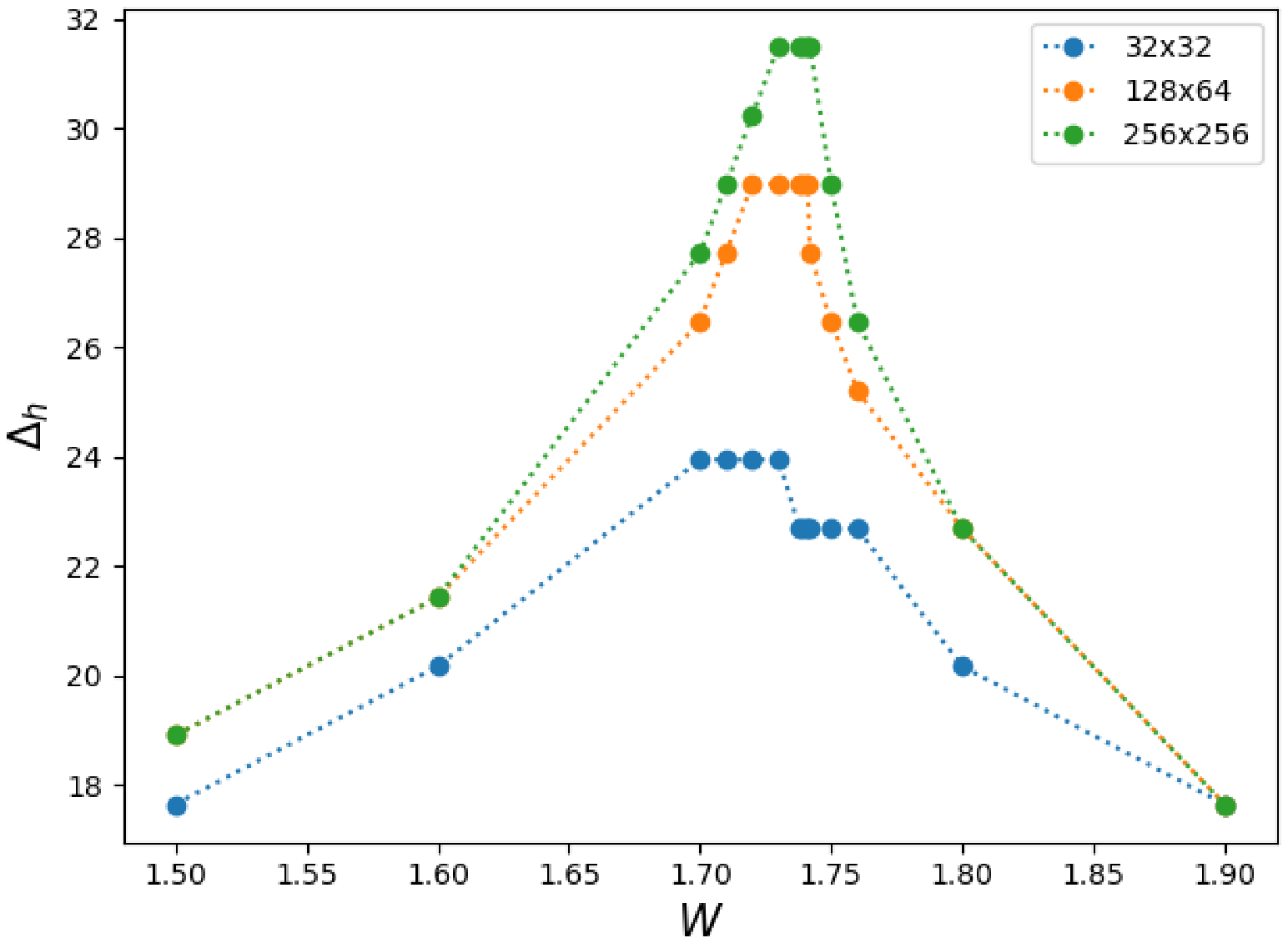}
        \label{fig:dcbc1}
    }

    \caption{a) First layer activity $\rho(r,W)$ as a function of stimulation
    rate $r$. The red line ($W_c = 1.74$) is the power law with exponent
    $m = 0.254$. Network size $N = 256 \times 256$. b) Dynamic range for the
    first layer, for several values of $W$ and systems with three different
    sizes.}
    \label{fig:resp1}
\end{figure}

As can be seen in Fig.~\ref{fig:respfl1}, the value of the exponent
$m \approx 0.254 \pm 0.005$ is  close to the expected value 
$m = \beta/\sigma = 0.268\pm 0.001 $, if we use the DP class  tabulated values
for $\beta$ and $\sigma$. Assis and Copelli~\cite{Assis2008} found comparable
values for the Stevens's exponent $m$ for a SIRS model in the square lattice.

As already observed, from our data we obtain
$\sigma = \beta/m = 0.63/0.254 = 2.48 \pm 0.05$. By using the tabulated
$\beta = 0.5834$, we get a better value $\sigma = 2.29 \pm 0.01$, to be compared
with the DP value  $\sigma = 2.1782 \pm 0.0171$~\cite{Lubeck2004}.

To understand how much the regimes influence the response consider
Fig.~\ref{fig:respfl1}. In the sub-critical regime, since the coupling between
neurons is small, the external input does not propagate and the system response
is linear. The same thing happens in the super-critical regime, but for a
different reason: in this case, we have self-sustained activity and small inputs
are lost in a noisy environment. It is only close to criticality (a few percent
from $W_c$) that both small and large stimulus alike can be mapped in the
output~\cite{Kinouchi2006,Assis2008,Shew2009,Larremore2011,Pei2012,Batista2014,Wang2017,Gautam2015}.
The dynamic range $\Delta_h$ of the first layer can be seem in
Fig.~\ref{fig:dcbc1}. At criticality, one can obtain a $\Delta_h \approx 32$ dB
for $L = 256$, and this value can be higher for larger networks.

\begin{figure}[hbt!]
    \centering
    \subfloat[][]{
        \includegraphics[width=0.49\linewidth]{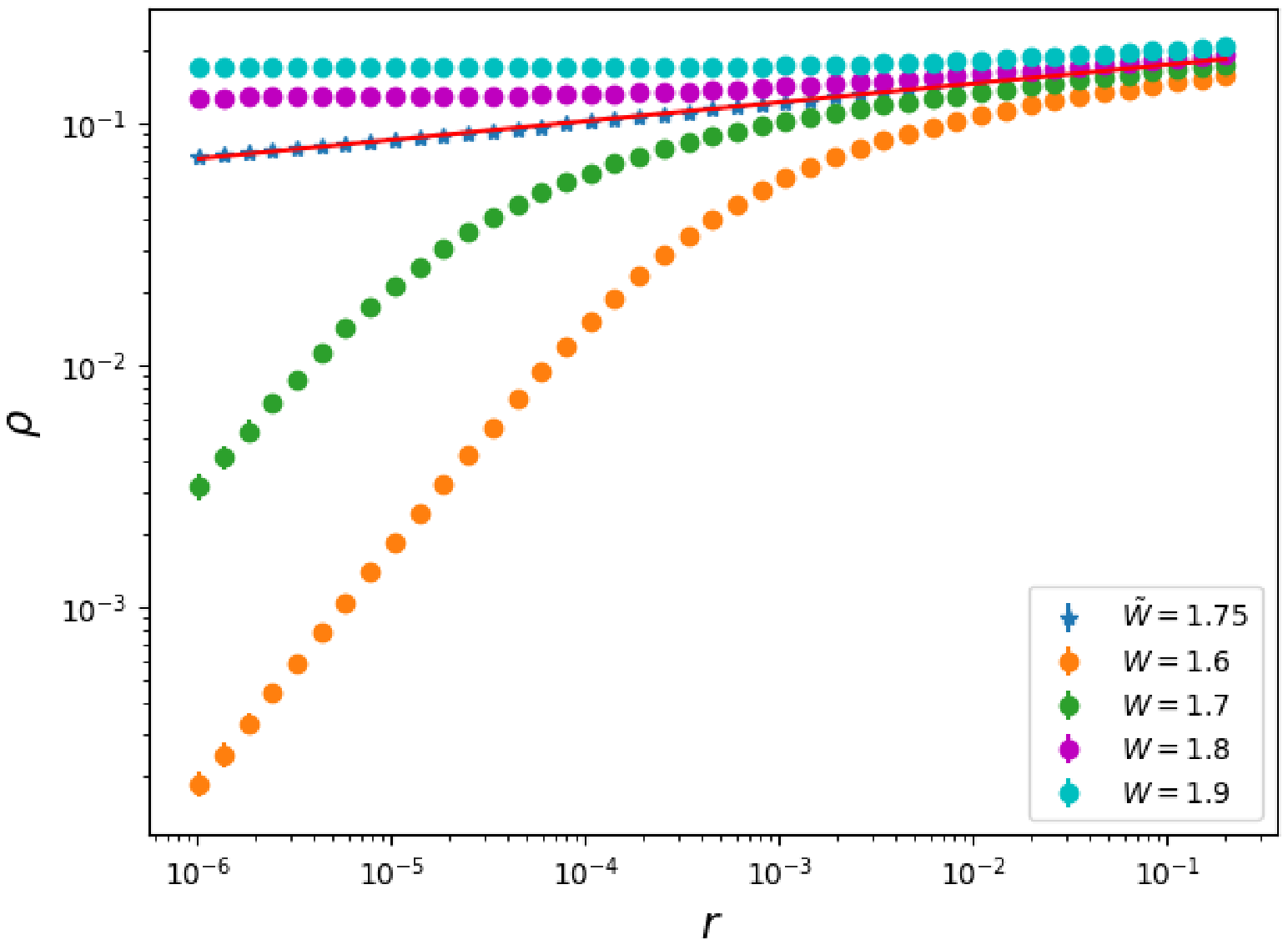}
        \label{fig:respfl2}
    }
    \subfloat[][]{
        \includegraphics[width=0.49\linewidth]{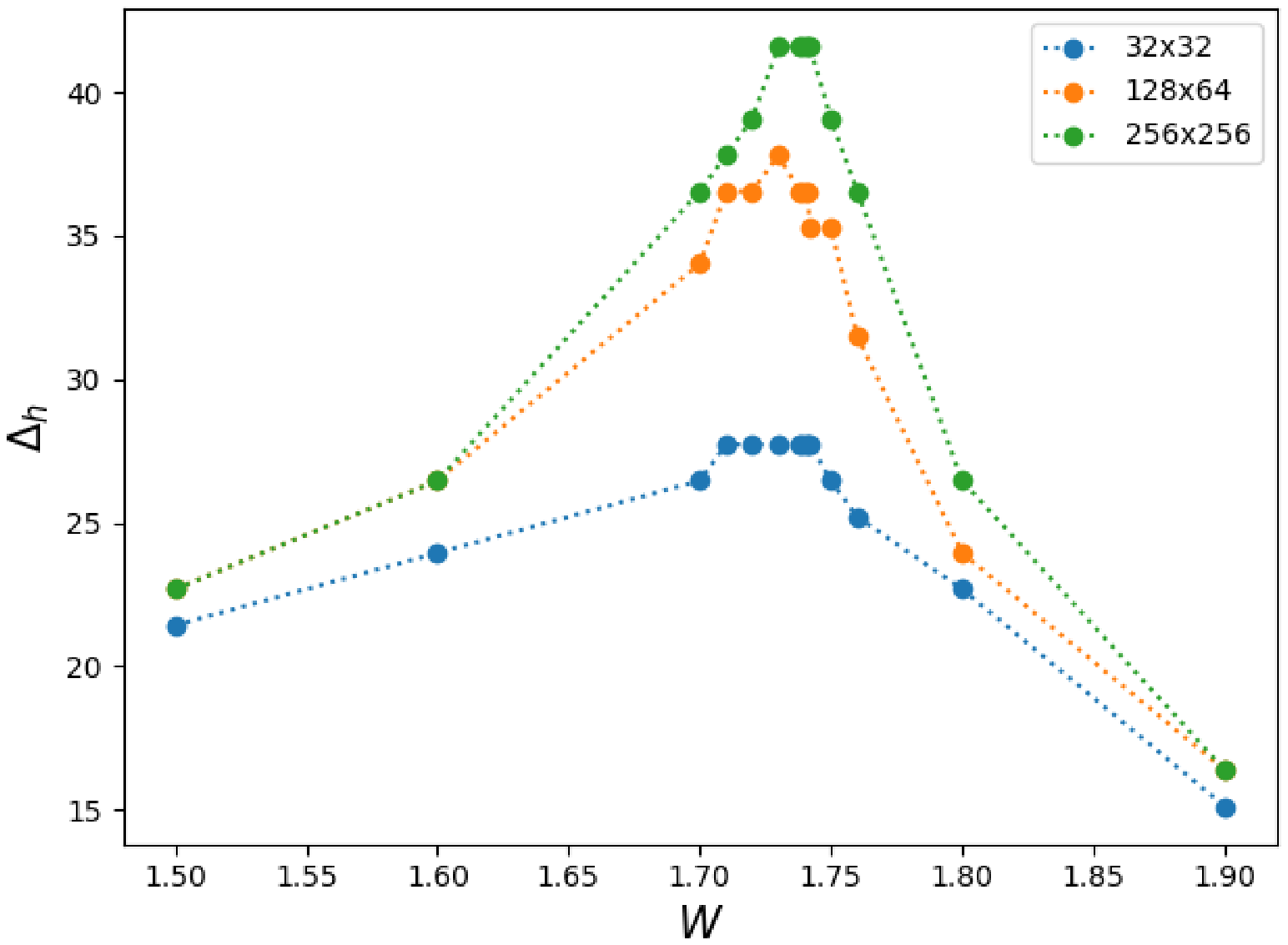}
        \label{fig:dcbc2}
    }

    \caption{a) Second layer neuronal activity $\rho_2(r,W)$ as a function of
    stimulation rate $r$ at the first layer. The red line ($W_c = 1.74$) refers
    to the power law Eq.~(\ref{eq:rhorl2}) with $m_2 = 0.078$. Network size
    $N = 256 \times 256$. b) Dynamic range for the second layer, for several
    values of $W$ and systems with three different sizes (the first layer has
    $N = 256 \times 256$ neurons).}
    \label{resp2}
\end{figure}

\subsection{Dynamic range of the second layer} 

Like the case of the first layer, we  present examples of the response function
$\rho_2$ in the three regimes, see Fig.~\ref{fig:respfl2}. The fit for the
critical power law gives $m_2 \approx 0.078 $ (the index refers to the second
layer). This accords with the expected value for  the exponent of
Eq.~(\ref{eq:rhorl2}), which is $m_2 = m^2 = 0.072$,  if we assume that the
first layer represents the external input for the second layer. Here, a fraction
$p=0.1$ of the neurons of the first layer are connected randomly to the second
layer by forcing synapses, that is, if the corresponding first neuron spikes,
the connected neuron in the second neuron spikes after a time step.

The dynamic range of the second layer in  the sub-critical and super-critical
regime are low for the same  reason they are small in the first layer. However,
in the critical regime, the dynamic range of the second layer is huge  (above
$40$ dB, see Fig.~\ref{fig:dcbc2}) and sufficient to account for the exquisite
performance of biological sensors.

\subsection{The effect of the inter-layer connectivity  $p$}

\begin{figure}[hbt!]
    \centering
    \includegraphics[width=0.49\linewidth]{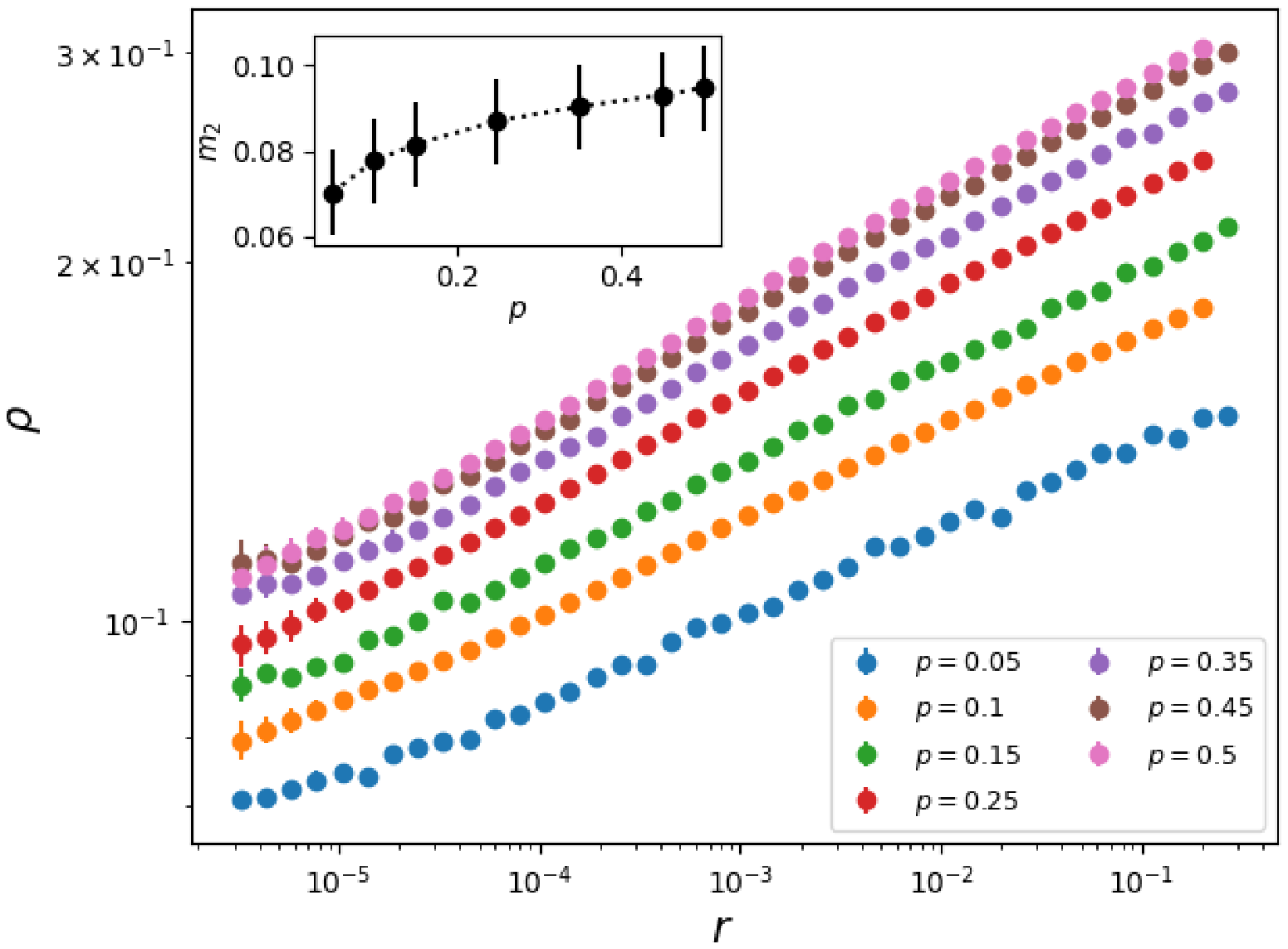}
     \includegraphics[width=0.49 \linewidth]{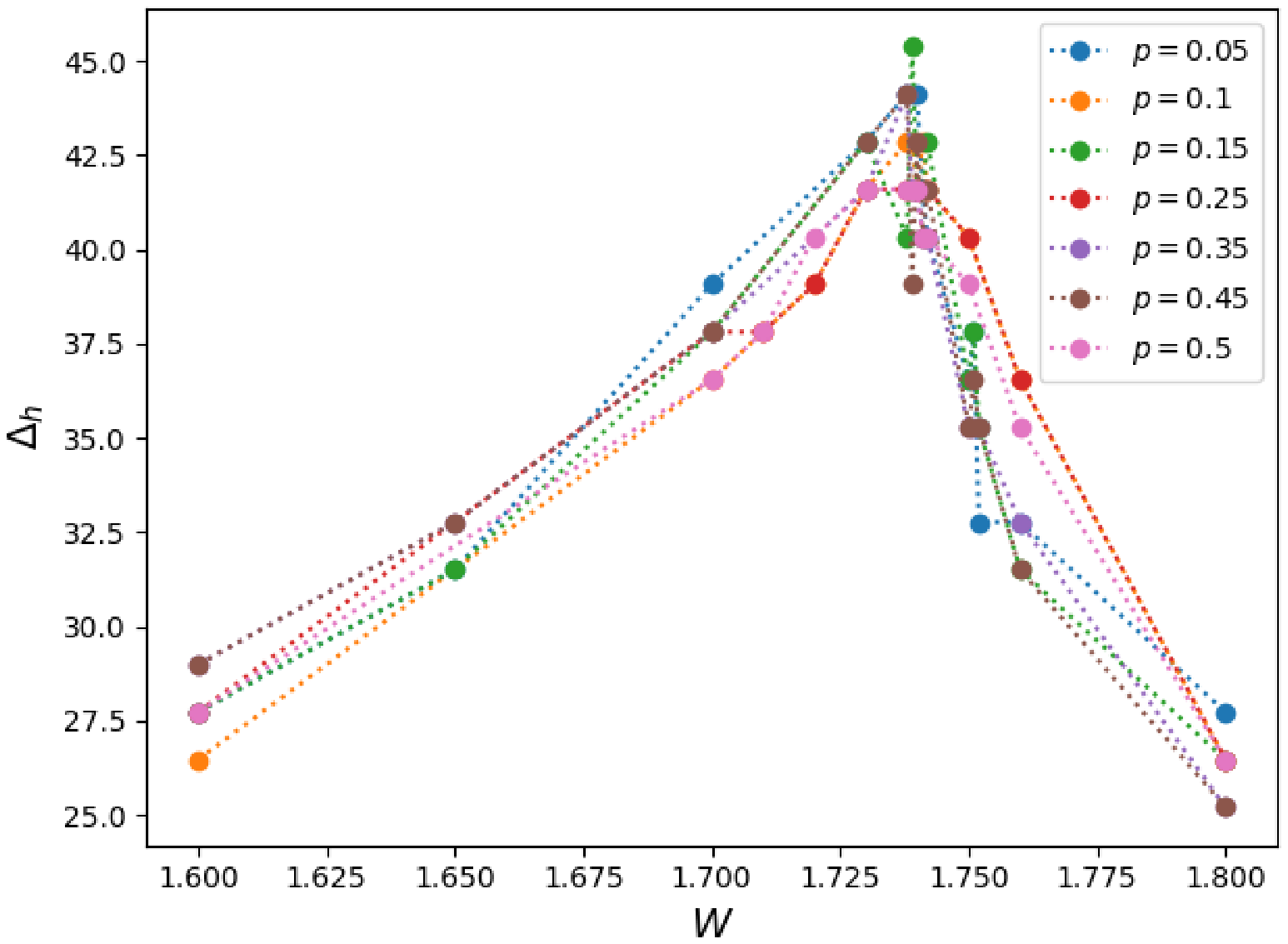}

    \caption{ a) Stimulus and response curves for various values of inter layer
    connectivity fractions $p$. The inset computes the exponent of the power law
    $\rho \propto r^{m_2}$ as a function of $p$. Simulation was carried using a
    $128\times128$ bi-dimensional network for each layer and fixed synaptic
    weight close to the critical point $W = 1.75$; b) Dynamic range for various
    values of $p$ for the second layer of a bi-layered system, each one consists
    of a $128\times128$ two dimensional network.}
    \label{fig:rhoxrpp}  
\end{figure}

As observed, for small interlayer connectivity $p=0.1$, the second layer
exponent seems to preserve the relation
$m_2 = 0.072 \approx m^2 = (1/\delta_h)^2$ \cite{Hinrichsen2000, Dickman1999}. 
However, for larger $p$, we deviate from this behavior, see
Fig.~\ref{fig:rhoxrpp} for simulations with $0.05 \leq p \leq 0.5$. 

We must remember here the origin of the increase of dynamic range in networks of
excitable cells (in contrast to pools of isolated cells in in recruitment
theory). Networks enable signal amplification, that is, the stimulation of one
cell by a input signal can produce a cascade of firings in neighbours  (a
branching process). This branching of the original signal means that small
signals are amplified, increasing the response to them. On the other hand,
the saturation due to very large input is delayed because the branching
processes interact and, since the cells have refractory periods, the activity is
suppressed~\citep{Copelli2002,Copelli2005,Copelli2005b}. This occurs both for
subcritical and suprecritical networks (their DR is always better than a pool 
with the same number of isolated neurons), but the optimal point is the critical
one~\cite{Kinouchi2006,Copelli2007}.

A very low $p$ means that the layers are uncoupled, so the signal amplification
mechanism does not works. On the other hand, a large connection $p$ means that
the activity of the first layer, that is already increased, is heavily
communicated to the second layer. Each neuron that receives a synapses is now
the source of a new branching process. This means that high $p$ induces
saturation in the second layer, increasing its Stevens exponent $m_2.$

Indeed, from Fig.~(\ref{fig:rhoxrpp})a, is possible to see that $m_2(p)$ is a
monotonically growing function of the connectivity $p$. This variation, however,
is not so large ($0.07 < m_2 < 0.10$), see Fig.~\ref{fig:rhoxrpp} inset. From
Fig.~(\ref{fig:rhoxrpp})b we see that the dynamic range for various $p$ is not
sensitive to such small variation.

\section{Homeostatic criticality}\label{H}

Up to now, we have shown that critical networks have maximal dynamic range.
However, we have not discussed how biological neuronal networks could tune
themselves towards criticality. In a series of 
publications~\cite{Campos2017,Brochini2016,Costa2017,Kinouchi2019,Girardi2020,Costa2015}
we have explored such homeostatic mechanisms, that implement the so called
Self-Organized quasi-Criticality (SOqC) scenario~\cite{Bonachela2009,Bonachela2010}. 

Homeostatic criticality means that the critical point turns out an attractor for
some adaptive dynamics of the system. In contrast to conservative SOC models
like sandpiles, where the self-organization depends on dissipation on the
borders of the system that have no explicit equations for that, SOqC models like 
forest fire models or neuronal networks have explicit drive and dissipation
mechanisms.

For example, lets consider activity dependent 
synapses~\cite{Levina2007,Campos2017,Girardi2020} such that they depress by a
factor $u$ when the presynaptic neuron fires (due to vesicle depletion) and
recover toward a baseline level $A$ with a  characteristic time $\tau$:

\begin{equation}
W_{ij,kl}[t+1] = W_{ij,kl}[t+1] + 
\frac{1}{\tau}\left(A - W_{ij,kl}[t] \right)
- u W_{ij,kl}[t] X_{ik}[t] \:\:\:\:\:\:\:\:\:\:\:\:\:\: 
(k \in {\cal V}) \:,
\end{equation}
where $k,l$ is the sites neighbors of neuron $i,j$. Here, the drive is the
recovering mechanism and the dissipation is due to the short term depression. It
is possible to show that, with this dynamics and for large $\tau$, the average
value  $W[t] = \avg{W_{ij,kl}[t]}$ goes towards $W_c$. 

We have noticed before that the second order phase transition only occurs  when
$\theta = I$, where $\theta = \avg{\theta_{ij}}$ and $I = \avg{I_{ij}}$ are the
average threshold and average input. This condition, for neuronal networks,
seems to be a fine tuning. However, if we think that the field $h = I - \theta$
is the average suprathreshold current, the condition of zero field $h=0$  is a 
natural requisite for continuous phase transitions in Statistical Physics.

So, inspired in firing rate adaptation mechanisms that postulate dynamic
thresholds~\cite{Benda2003}, we propose the following homeostatic dynamics:

\begin{equation}
    \theta_{ij}[t+1] = \theta_{ij}[t] -\frac{1}{\tau_\theta}
    \theta_{ij}[t] + u_\theta \theta_{ij}[t] X_{ij}[t] \:,
\end{equation}
where now the signs are inverted: the dissipation is due to the $1/\tau$ decay
and the drive (grow of the threshold) occurs when   the neuron  spikes. It is
also possible to show that this adaptive dynamics leads to
$\theta[t] = \avg{\theta_{ij}[t]}  \rightarrow I$, that is, $h \rightarrow 0$.
An experimental prediction of this mechanism is that, in critical neuronal
networks with power law avalanches, the neurons mostly adapt their firing 
thresholds to their external inputs.

So, with these two homeostatic mechanism, the critical point $W=W_c,h=h_c=0$ in 
the phase diagram of the systems turns out an attractor of the overall dynamics.
Simulations of these homeostatic mechanisms in the 2d lattice are somewhat out
of the scope of this paper, but full results will be presented in future work.

\section{Discussion and Conclusion}

The biological motivation for our the model is the retina, where  both lateral
and vertical coupling by electric synapses (gap junctions) occur, forming
neuronal networks with stacked layers~\cite{Cook1995}. All these electric
synapses are plastic, from the milisseconds to the minutes time
scales~\cite{OBrien2014}, what opens the possibility of homeostatic tuning to
criticality~\cite{Kinouchi2019,Girardi2020} as supposed here. Moreover, there is
experimental~\cite{Deans2002,Murphy2011} and computational~\cite{Publio2009}
evidence that disruption of electric synapses diminishes the sensitivity  and
degrades the retina dynamic range. We emphasize that we worked here with
stochastic integrate-and-fire neurons, not cellular automata as done
in~\cite{Kinouchi2006,Copelli2007, Assis2008,Campos2017,Copelli2005,Gollo2009,Martinez2020}, 
generalizing thus these results to biologically more realistic elements.

By studying the critical exponents at the second order phase transition, we
found that 2d lattices of stochastic integrate-and-fire neurons are compatible
with the Directed Percolation universality class. We then proposed the topology
of two coupled square lattices to increase the dynamic range of a retina-like
sensor.  The first one receives Poisson inputs at rate $r$, and represents it as
a neuronal activity $\rho_1 \propto r^m$, with
$m = 1/\delta_h = \beta/\sigma = 0.268$. This activity is passed, by a fraction
$p$ of neurons, to the second layer which then presents an output activity 
$\rho_2 \approx r^{m_2}$. The final Stevens's exponent of the system is
$m_2 = 0.078 \approx m^2 = (\beta/\sigma)^2 = 0.072$. Thus, the exponent
relation Eq.~(\ref{eq:meq}) proposed in \cite{Kinouchi2006} seems to be valid,
regardless of topology, as long as the stimulus intensity is moderate: the power
law response is valid only before a saturating regime (Hill's like curve) also
found in biological sensors.

It is possible to show that 1d systems (a ring of neurons) pertain  to the 1d DP
class~\cite{Assis2008,Pazinni2020} (or perhaps the 1d Manna class). In this
case, we have a very large dynamic range due to the expected value
$m = \beta/\sigma =  0.276486/$ $ 2.554216 = 0.108247$. This means that an input
range of ${\cal O}(10^{15})$ units can be mapped to an output range of
${\cal O}(100)$. Although such low dimensional topologies perhaps have no
applications in Biology, it is conceivable that artificial sensors with huge
dynamic range could be constructed based in these principles.

\section{Acknowledgement}

EG thanks CAPES for  financial support. OK acknownledges CNAIPS-USP support and
FAPESP scholarship BPE 2019/12746-3. This work was produced as part of the
activity of FAPESP Research, Innovation and Dissemination Center for
Neuromathematics (grant \#2013/07699-0 S. Paulo Research Foundation). The
present work was also realized with the support of CNPq, Conselho Nacional de
Desenvolvimento Científico  e Tecnológico - Brasil.

\bibliographystyle{unsrt}
\bibliography{Bib}

\end{document}